\documentstyle{l-aa}

\newcommand{\bc}{\begin{center}}      
\newcommand{\ec}{\end{center}}
\newcommand{\be}{\begin{equation}}    
\newcommand{\ee}{\end{equation}}

\newcommand{\kms}{\mbox{km s$^{-1}$}}

\begin{document}

   \thesaurus{03         
              (11.19.2;  
               11.05.2;  
               11.11.1;  
               11.19.6;  
               11.09.1;  
               03.13.4)} 

\title{Gasdynamics in NGC 4736}

\author{Qiu-Sheng Gu\inst{1} \and Xin-Hao Liao\inst{1}
  \and Jie-Hao Huang\inst{1} \and Qin-Yue Qu\inst{1} \and Hongjun Su\inst{2}}

\offprints{Q.-S Gu, e-mail:postcstd@nju.edu.cn}

\institute{Department of Astronomy, Nanjing University, Nanjing 210093, China
\and Purple Mountain Observatory, Nanjing 210008, China}

\date{Received ; accepted }

\maketitle

\begin{abstract}

 New test-particle simulations have been performed to study
 the secular evolution of the gaseous distribution in NGC 4736. We find 
 that the distribution of gas clouds can be understood in the frame of
 perturbation induced by a nuclear oval potential, in addition to a
 spiral-like potential. Our experiments show
 that both inner and outer rings are stable structures located at the inner
 Lindblad resonance (ILR) and the outer Lindblad resonance(OLR), 
 respectively, in agreement with the observations of NGC 4736.
  One of our simplified simulations indicates that
 both nuclear starburst and orbital resonance might be needed for 
 keeping the inner ring stable for a long period.
 We have introduced a  symplectic algorithm
 in the orbit integration. Its stronger stability allows us to adopt a
 rather large time step to save computational time. Substitution 
 of a viscous force for cloud-cloud interactions proves adequate in the
 case of NGC 4736.

\end{abstract}

\section{Introduction}

NGC 4736 (M94) is a bright, nearby Sab galaxy \footnote{The Hubble distance 
is still uncertain, we adopt 6.3 Mpc in this paper, so that 1 arcmin is 
equal to 1.8 kpc.}, which is notable for a faint
outer H\,{\sc i} ring of radius 4 arcmin to 6 arcmin, and an inner bright ring
of H\,{\sc ii} region with enhanced gas density at 45 arcsec (Lynds  1974;
Beckman et al. 1991). 
Beckman et al. (1991) found that the bulge position angle ($\sim 20^\circ $) 
was very different from that of the disc ($\sim 120^\circ $), 
strong evidence for non-axisymmetry in the bulge of NGC 4736, and suggested that
triaxiality of the bulge played an important dynamical role in fuelling
star formation.
 
Discussion of the origin of these rings is separated into two groups.
On one hand, van der Kruit(1974, 1976) and Sanders \& Bania(1976)
suggested that the ring-like structures in NGC 4736 are the observational
signature of recent nuclear explosive events.
On the other hand, Schommer \& Sullivan (1976) and Bosma et al. (1977)
proposed that the rings result from tight winding of spiral arms
and are tracers for the principle orbital resonance regions. Athanassoula 
et al. ( 1982 ) thought that NGC 4736 is in fact the prototype of rings located
at Lindblad resonances in an oval potential.

Gerin, Casoli \& Combes (1991) have shown that both inner H\,{\sc ii} and
outer H\,{\sc i} rings could form in a slight barred potential. 
The possible existence of such an oval was first discussed by Bosma, van
 der Hulst \& Sullivan (1977) and Kormendy (1979) and more recently by 
 Huang, Gu \& Su (1993). As pointed out by the authors themselves, the
 model of Gerin, Casoli \& Combes has two main drawbacks. Their outer ring
 is too wide (cf. their Fig. 10b), so that it was in fact associated
 with both corotation at about 4.5 kpc and OLR at about 8.0 kpc. The second 
 drawback is that their inner ring disappears after 3 Gyrs. The aim of this 
 paper will be to produce a model devoid of these drawbacks.

Here, we present our numerical experiments on gas clouds in NGC 4736 based on 
the test-particle method. We replace the process of cloud-cloud collision
with the viscous force widely used in the study of accretion
disks and introduce the symplectic algorithm, which has proven of strong stability, 
in our simulations. Both inner and outer ring structures turned
out to be quite stable after 4.0 Gyr, and are associated with the inner
Lindblad resonance (ILR) and outer Lindblad resonance(OLR) for a long duration, respectively.
 
In Sect.2, we describe  the construction of the model potential of NGC 4736, and 
the method of replacing the process of collision between gas clouds.
The computational method, namely, the near-symplectic algorithm, is described 
in Sect.3, and in Sect.4 we present the simulation results of the gasdynamics 
in NGC 4736. Finally, we give the discussions and conclusions in Sect.5 and 6.

\section{The new simulations}

We assume in our new simulations that the motions of the gas clouds are mainly 
dominated by the following two physical processes: the galaxy's gravitational 
field and the mutual collision between gas clouds. Note that  we neglect
the self-gravity of the gas cloud, which may have an important effect 
in some cases.

\subsection{The gravitational field}
Following the basic approach, 
the galactic potential consists of two parts, an axisymmetric background 
and a non-axisymmetric perturbing potential. 
We assume that the background potential has
four components: a spherical bulge, a thick disk, a thin disk, and a 
spherical halo. 
Both the density distribution of the bulge and of the halo take the form 
\begin{equation}
 \rho = \left \{ \begin{array}{ll}
   \frac{\rho_{0}}{2\pi}(1-\frac{r}{r_{\rm c}}),   &  r \leq r_{\rm c}, \\
      0 , & r > r_{\rm c}.
  \end{array} \right.
\end{equation}
\noindent
which implies an underlying bulge gravitational 
potential of the form
\begin{equation}
U_{\rm B} = \left \{ \begin{array}{ll}
            -2.0 G \rho_{0} r^{2} (\frac{1}{3}-\frac{r}{4r_{\rm c}}) , &  r \leq r_{\rm c}, \\
            -G \rho_{0} r_{\rm c}^{3}/6r                             , &  r > r_{\rm c}.
  \end{array} \right.
\end{equation}
\noindent
The halo potential $U_{\rm H}$ has the same form as $U_{\rm B}$. The thick
and thin disks
are assumed to be Toomre disks (Toomre, 1963), and have the form 
\be
 U_{\rm D}(r)=-\frac{GM_{\rm D}}{(r_{\rm c}^2+r^2)^{1/2}},
\ee
\noindent

We take the perturbing potential as the same as that used in Roberts \& 
Hausman's 
 simulations(1984).
\be
U_{1}(r,\theta,t)=U_{\rm D}(r) \frac{A}{5}\frac{r_{\rm c}^2r^2}{(r_{\rm c}^2+r^2)^2} 
\cos[2\theta-2\Omega_{\rm p}t+\Phi(r)].
\ee
where $\Phi(r)$, the phase of the maximum perturbation, is determined by
\be
\Phi(r) = 2 \log (1+(r/r_{0})^{j})/(j \tan i_{0}),
\ee
\noindent
In the above equations, $r_{\rm c}$ is a constant canonical radius,
$\Omega_{\rm p}$ is pattern speed, $M_{\rm D}$ the 
total mass of the Toomre disk, $r_{0}$ a characteristic radius where 
the perturbing potential changes from barlike to spiral-like. The 
power $j$ determines how sharply the shift from barlike to spiral 
perturbation occurs with increasing $r$, and $i_{0}$ is the pitch 
angle of the spiral. In our model, $r_{0}$, $j$, 
and $i_{0}$ are set to be 1.0kpc, 5, and 10$^{o}$, respectively. 

\begin{figure}
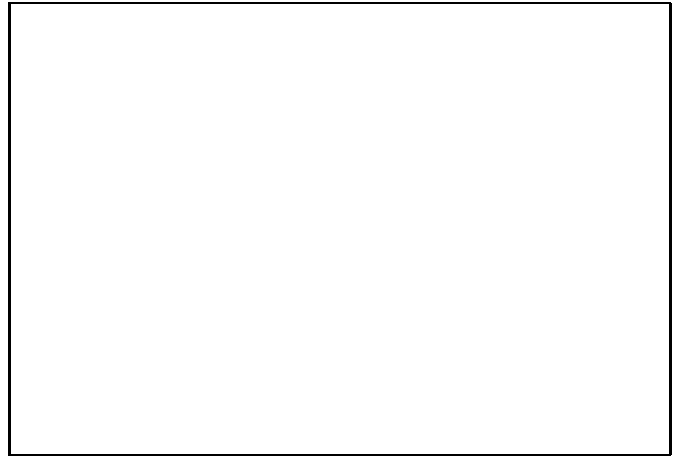

\picplace{6cm}
\caption{Rotation curve model of NGC 4736}
\end{figure}

\begin{figure}
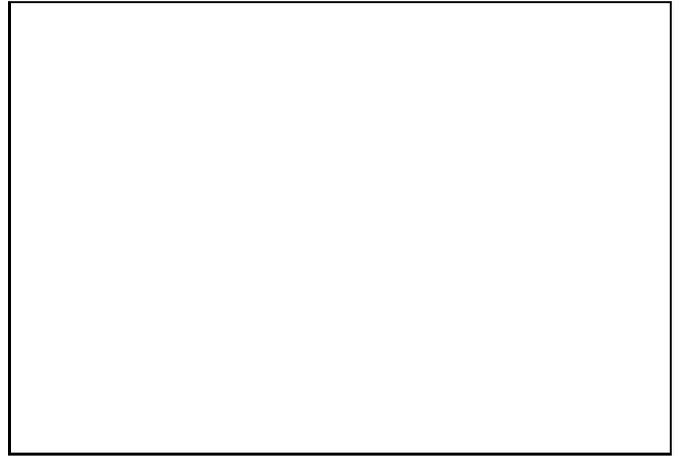

\picplace{6cm}
\caption{Circular angular velocity ( $\Omega$ ) and Lindblad precession frequencies
( $\Omega \pm \kappa/2$ ) vs. radius for the rotation curve model of NGC 4736}
\end{figure}

Now we can construct a mass model of NGC 4736 by fitting the observed 
rotation curve with the model potentials mentioned above. The rotation
curve of NGC 4736 has the following characteristics:
a steep rise from the nucleus to 0.25 arcmin, a turn-over region from 0.25 
to 0.5 arcmin, a constant-velocity region from 0.5 to 4.0 
arcmin and a decreasing-velocity region from 4.0 to 6.0 arcmin 
(Schommer \& Sullivan 1976 ; Bosma et al. 1977). 
The mass parameters, determined this way, of the bulge, thick disk, thin disk
and halo are 0.005, 0.032,  0.11, and  0.11 $kpc^{3}/Myr^{2}$, 
respectively, and the constant canonical radii for these four components
are 0.30, 0.45, 2.50, and 8.20 kpc, respectively.  Fig.1 gives the 
rotation curve model corresponding to the unperturbed gravitational field defined 
by the above parameter values, which is almost in keeping with that derived 
from optical and radio observations. The standard Lindblad precession frequencies for 
this rotation curve are shown in Fig. 2.

The equation of motion of gas clouds in the potential $U(x,y,t)$ can be 
written as
\be
  \left \{ \begin{array}{cc}
   \frac{dx}{dt} = \frac{\partial H}{\partial \dot{x}}, & \frac{d\dot{x}}{dt} =  - \frac{\partial H}{\partial x},\\
   \frac{dy}{dt} = \frac{\partial H}{\partial \dot{y}}, & \frac{d\dot{y}}{dt} =  - \frac{\partial H}{\partial y},
  \end{array} \right. 
\ee
\noindent
where, $H$, the Hamiltonian of the gas cloud, is represented by
\be
 H=H(x,y,\dot{x},\dot{y},t)= \frac{1}{2}(\dot{x}^2 + \dot{y}^2) + U(x,y,t),
\ee
\noindent
$\dot{x}$ and $\dot{y}$ are the canonical conjugate variables of $x$ and $y$, 
respectively.

\subsection{The mutual collisions between gas clouds}
The cloud-cloud collision is obviously related to the following parameters: 
the gas density, the velocity dispersion, the collision cross-section, the 
constitution of matter, the differential rotation,
the included collision angle, etc. Therefore, it is 
difficult to formulate the collision processes exactly in numerical simulations. 
The effect of the mutual collision between gas clouds was treated as a simple 
fractional reduction of the relative velocities first by Schwarz (1981,1984).
In our present numerical
experiments, we introduce a dissipative viscous force to simulate the effect
of collision, expressed by the following formula in the well-known 
$\alpha$-disk (Shakura and Sunyaev 1973; Pringle 1981),
\be
 F = \alpha C_{\rm s} \Sigma h r \frac{d\Omega}{dr},  
\ee
\noindent
where, $\alpha$ is a constant and $ 0<\alpha<1 $; 
$C_{\rm s}$ the local sound velocity; $\Sigma$ the cloud surface density; 
$\Omega$ the local angular velocity; and $h$ the spiral thickness.
For the gas clouds in a galaxy, $C_{\rm s}$
is just the local velocity dispersion. As a first-order approximation, we 
take $\vec{v}-\vec{v}_{0}$ as $C_{\rm s}$, where $\vec{v}$ and $\vec{v}_{0}$ is 
the local velocity and the local unperturbed circular rotation speed, 
the initial number density of gas clouds $\Sigma_{0}$ as $\Sigma$, 
and the local  unperturbed circular rotation $\Omega_{0}$ as $\Omega$. 
So, we have 
\be
\vec{F}= \alpha (\vec{v}-\vec{v}_{0}) \Sigma_{0} h r \frac{d\Omega_{0}}{dr}
\ee
\noindent
to replace the effect of collision between gas clouds.
Obviously, $\vec{F}$ is a small quantity of first-order magnitude 
as compared with the 
unperturbed gravitational force. With the addition of $\vec{F}$, the motion 
equation of a single gas cloud on the circle of radius $r$ becomes 

\be
  \left \{ \begin{array}{cc}
   \frac{dx}{dt} = \frac{\partial H}{\partial \dot{x}}, & \frac{d\dot{x}}{dt} =  - \frac{\partial H}{\partial x} + 2\pi rF_{\rm x},\\
   \frac{dy}{dt} = \frac{\partial H}{\partial \dot{y}}, & \frac{d\dot{y}}{dt} =  - \frac{\partial H}{\partial y} + 2\pi r F_{\rm y},
  \end{array} \right. 
\ee
\noindent
where $F_{\rm x}$ and $F_{\rm y}$ are the components of $\vec{F}$ in the 
directions of $x$ and $y$, respectively.

\section{The computational method}

 For the numerical computation of the Hamiltonian system, we could adopt 
traditional methods such as Runge-Kutta. But most of the 
traditional numerical methods will cause a linear variation of the 
system energy with time, contrary to the theoretical results for Hamiltonian 
systems. It is known that the Hamiltonian flow preserves its symplectic
structure (Arnold, 1978). Based on this property, a symplectic algorithm
for Hamiltonian systems 
has been developed (Ruth 1983; Feng 1984). Its 
obvious advantage over the traditional methods is in the aspect of the long-term 
dynamical evolution of Hamiltonian system, i.e. no secular change in the system energy.

For the Hamiltonian system

\be
 \left \{ \begin{array}{lr}
 \dot{q} = & \frac{\partial H}{\partial p}, \\ 
 \dot{p} = & -\frac{\partial H}{\partial q}, 
 \end{array} \right.
\ee
\noindent
with Hamiltonian function
\be
 H = H(p,q) = T(p) + V(q),
\ee
\noindent
where $p, q \in R^{n}$ are a set of canonical conjugate variables, its phase 
flow is represented by
\be
 \left ( \begin{array}{c}
  p(t) \\ q(t)
 \end{array} \right )
 = exp[(t-t_{0})D_{\rm H}]
 \left ( \begin{array}{c}
  p(t_{0}) \\ q(t_{0})
 \end{array} \right ),
\ee
\noindent
where the differential operator $D_{\rm H}$ is defined by the Poisson bracket 
$\{\cdot,\cdot\}$ as follows
\be
  D_{\rm H} = \{\cdot , H\}
\ee
\noindent
and $exp[(t-t_{0})D_{\rm H}]$ is the exponential transformation of 
$D_{\rm H}$. Let the integration time-step be $\tau$, both of 
the following step-transition operators
\be
 g_{1}^{\tau} = exp(\tau D_{\rm T})exp(\tau D_{\rm V})
\ee
\be
\breve{g}_{1}^{\tau} = exp(\tau D_{\rm V})exp(\tau D_{\rm T})
\ee
\noindent
are first-order symplectic integrators. It is easy to prove that both
\be
 g_{2}^{\tau} = g_{1}^{\tau} \circ \breve{g}_{1}^{\tau}
\ee
\noindent
and
\be
 \breve{g}_{2}^{\tau} = \breve{g}_{1}^{\tau} \circ g_{1}^{\tau}
\ee
\noindent
are second-order symplectic algorithms. Higher-order symplectic algorithms 
can be constructed from a certain number of second-order symplectic 
algorithms (Yoshida, 1990).

The difference scheme corresponding to $g_{2}^{\tau}$ is
\be
 \left \{ \begin{array}{lll}
   q_{\rm k+\frac{1}{2}} & = & q_{\rm k}+\frac{\tau}{2}\frac{\partial T(p)}{\partial p} |_{p=p_{\rm k}}, \\
   p_{\rm k+1} & = & p_{\rm k} - \tau \frac{\partial V(q)}{\partial q} |_{q=q_{\rm k+\frac{1}{2}}}, \\
   q_{\rm k+1} & = & q_{\rm k+\frac{1}{2}}+\frac{\tau}{2}\frac{\partial T(p)}{\partial p} |_{p=p_{\rm k+1}},
 \end{array} \right. 
\ee
\noindent
where $(p_{\rm k+1}, q_{\rm k+1})$ and $(p_{\rm k}, q_{\rm k})$ are the
numerical 
solutions of the Hamiltonian flow (13) at time $t = (k+1)\tau$ and $t=k\tau$, 
respectively, and  $p_{0}=p(t_{0})$, $ q_{0}=q(t_{0})$.

\begin{figure*}
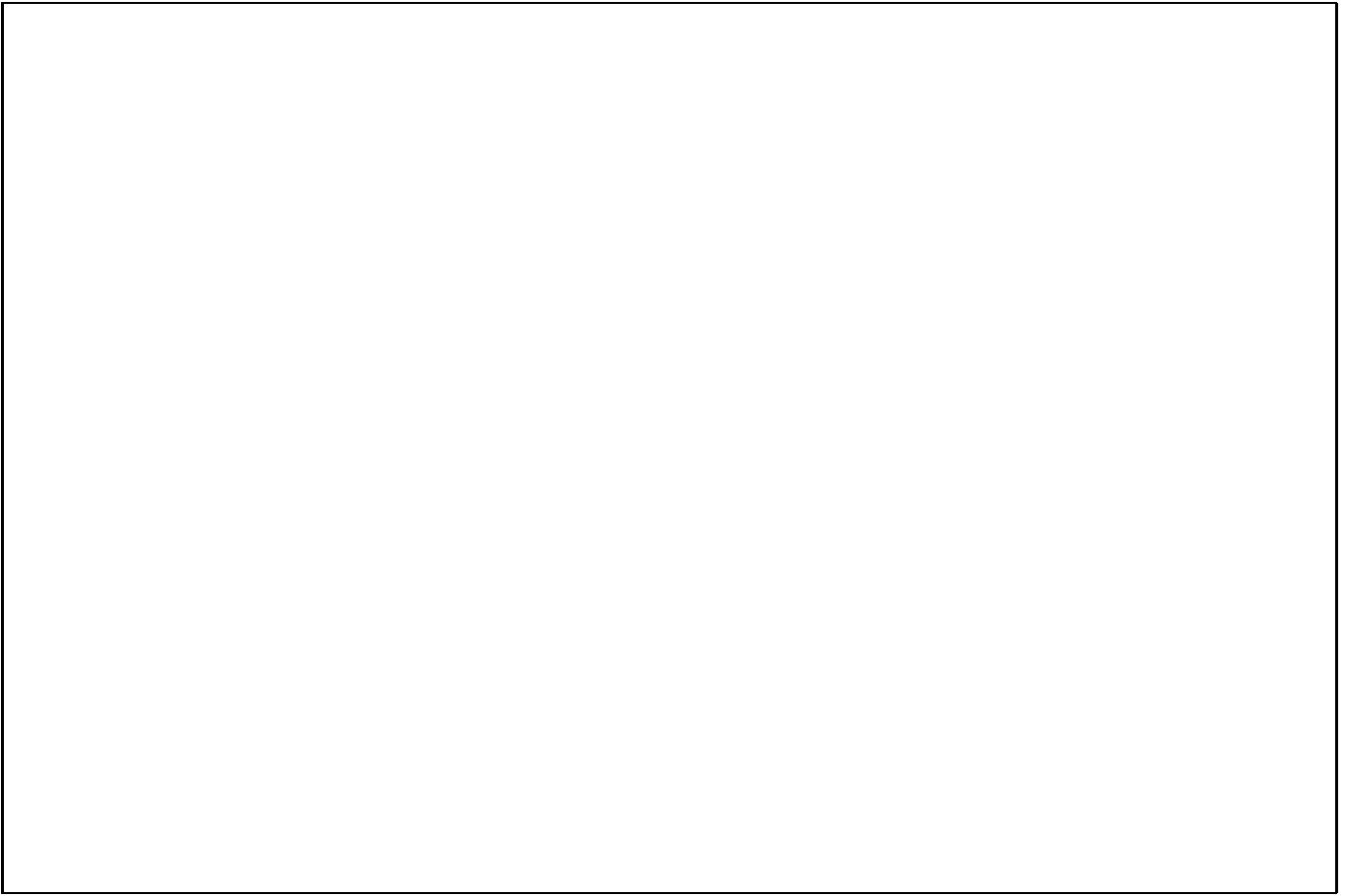

\picplace{12cm}
\caption{Time evolution of the gas clouds in our model. The time epochs for 
a, b, c, d, e, and f are 50 Myr, 200 Myr, 400 Myr, 1.0 Gyr, 2.0 Gyr and 
4.0 Gyr, respectively.}
\end{figure*}

If we consider the small dissipative effect, then equation (11) becomes
\be
 \left \{ \begin{array}{ll}
 \dot{q} = & \frac{\partial H}{\partial p}, \\
 \dot{p} = & -\frac{\partial H}{\partial q} + \varepsilon F(p,q) ,
 \end{array} \right.
\ee
\noindent
Since $\varepsilon$ is a small quantity, we could replace the 
solution of the second equation (20) by an Euler flow. Hence the difference 
scheme (19) changes to
\be
 \left\{ \begin{array}{lll}
   q_{\rm k+\frac{1}{2}} & = & q_{\rm k}+\frac{\tau}{2}\frac{\partial T(p)}{\partial p} |_{p=p_{\rm k}}, \\
   p_{\rm k+1} & = & p_{\rm k} - \tau [\frac{\partial V(q)}{\partial q} + \varepsilon F(p,q)] |_{p=p_{\rm k}, q=q_{\rm k+\frac{1}{2}}}, \\
   q_{\rm k+1} & = & q_{\rm k+\frac{1}{2}}+\frac{\tau}{2}\frac{\partial T(p)}{\partial p} |_{p=p_{\rm k+1}},
 \end{array} \right. 
\ee
\noindent
This kind of treatment causes the truncation error of (21) to be of order 
max$[O(\tau^{3}), O(\tau^{2}\varepsilon)]$. Obviously, 
we can ensure that the scheme (21) is second-order if $\varepsilon < \tau$,
and we 
will use this scheme in our simulation.

  It should be noticed that the ``leap-frog" difference scheme used 
in the $N$-body simulation is also symplectic and second-order. But 
its stable interval 
is very small in comparison with scheme (21).

\section{Results}

At the beginning of each simulation, 10000 gas clouds are
uniformly distributed in the disk with 10 kpc radius. Each cloud is 
given with a local circular rotation velocity, along with a peculiar 
velocity selected from a two-dimensional Gaussian distribution whose 
one-dimensional dispersion is 5 $\kms$.

\begin{figure*}
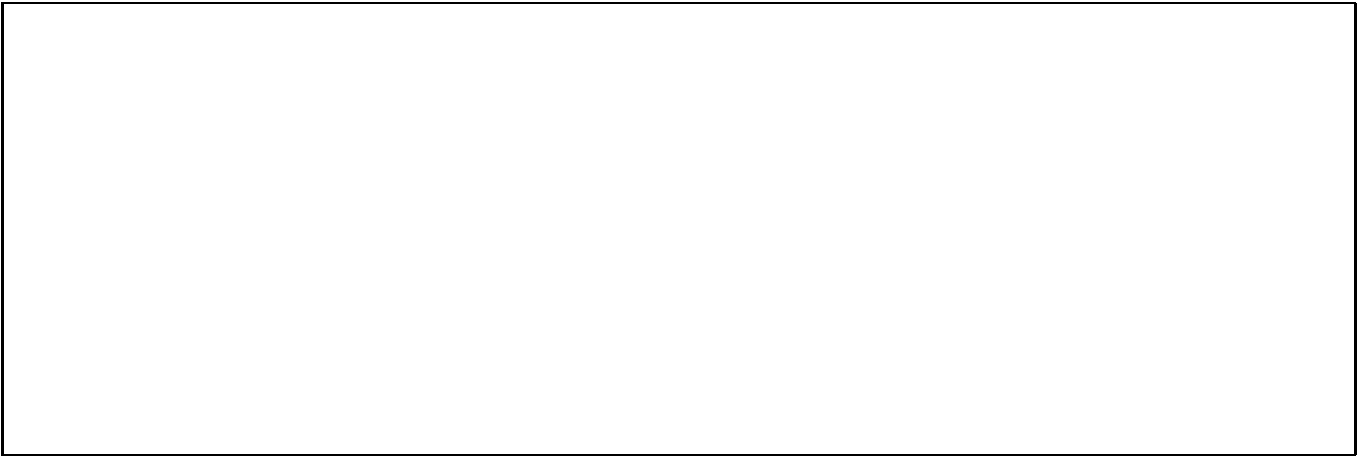

\picplace{6cm}
\caption{Time evolution of the gas clouds, to which within the ILR region some
radial motions are added when $\Sigma > \Sigma_{c}$.
The time epochs for a, b and c are 500 Myr, 1.0 Gyr, and 3.6 Gyr, respectively.}
\end{figure*}

With the amplitude of the spiral perturbation $A=0.4$ , the 
pattern speed $\Omega_{\rm p} = 0.03 $, and the viscous coefficient 
$\alpha=0.001$, the 
dynamical evolution of gas clouds with time is given 
in Fig.3. It is clearly shown that when time $ t = 50 Myr $, Fig. 3a, a faint
ring-like structure of radius 1.1kpc emerges in the central region; with 
increasing time, more and more gas particles move inward
and accumulate at the ring region. At $ t = 400 Myr$, the inner 
stable ring becomes prominent, roughly located at the ILR with a radius 
of 1.1 kpc , see Fig. 3c.  At the same time, in the outer part of the disk, 
an outer ring is formed as the result of tight winding of spiral arms.
It is stable and of low-density with radius 
7.5 kpc to 10.0 kpc, a location near the OLR and beyond. It is interesting
to notice that both inner and outer ring structures are stable for 
durations of 100 Myr to 800 Myr, fixed onto the ILR and OLR, respectively, 
in agreement with the observations of NGC 4736.

After 1.0 Gyr, the inner ring moves inward further, to a radius of 
about 0.7kpc, see Fig. 3d, and stays there till 4.0 Gyr, see Fig. 3f. 
Meanwhile, the outer ring appears narrower, but remains in the OLR region.

\begin{figure*}
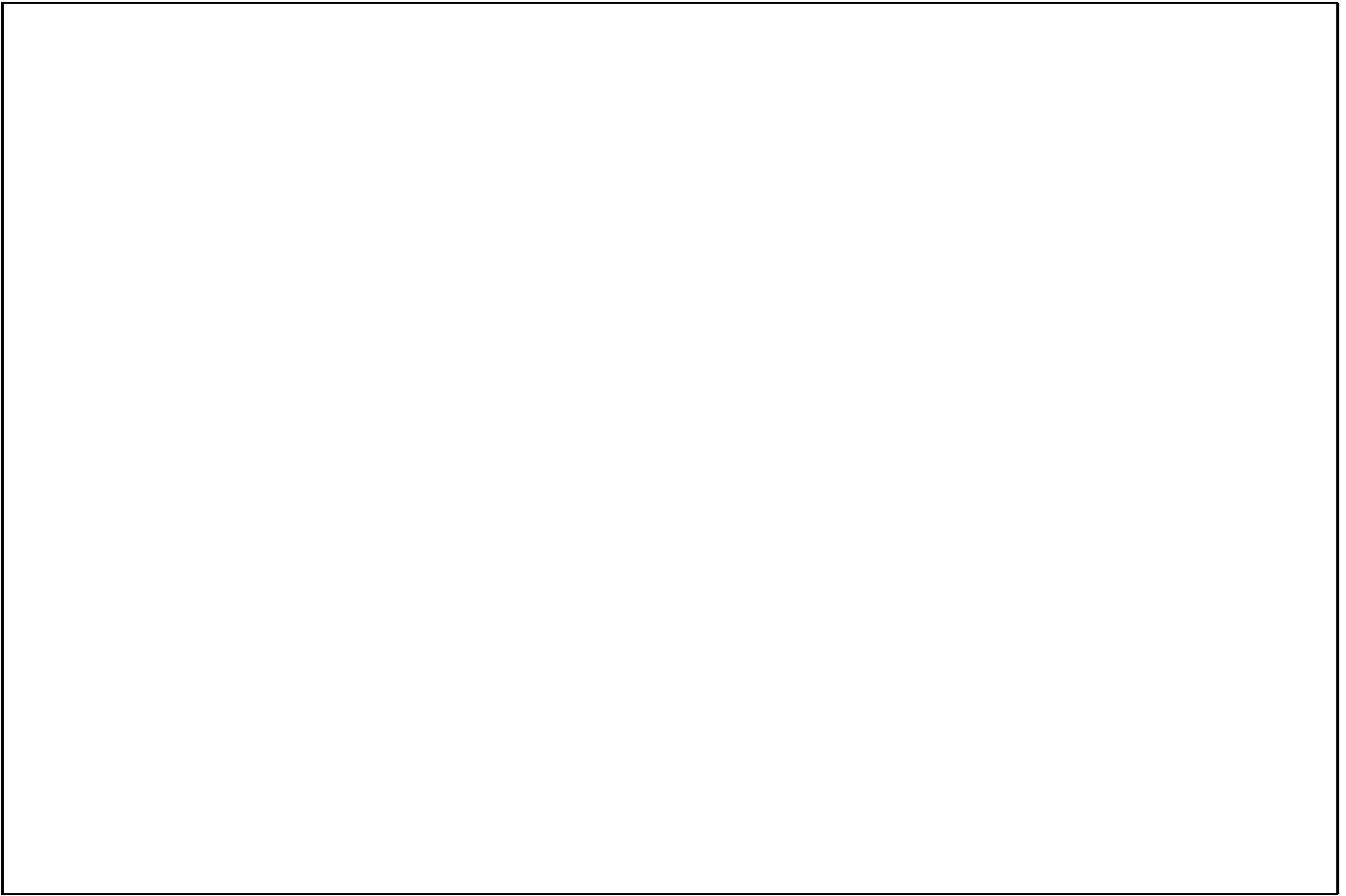

\picplace{12cm}
\caption{Time evolution of the gas clouds, using Schwarz's potential but our
model for cloud-cloud interactions. The time epochs in bar rotations are
2, 4, 6, 8, 10, and 20 for frames a, b, c, d, e, and f, respectively.}
\end{figure*}

\begin{figure*}
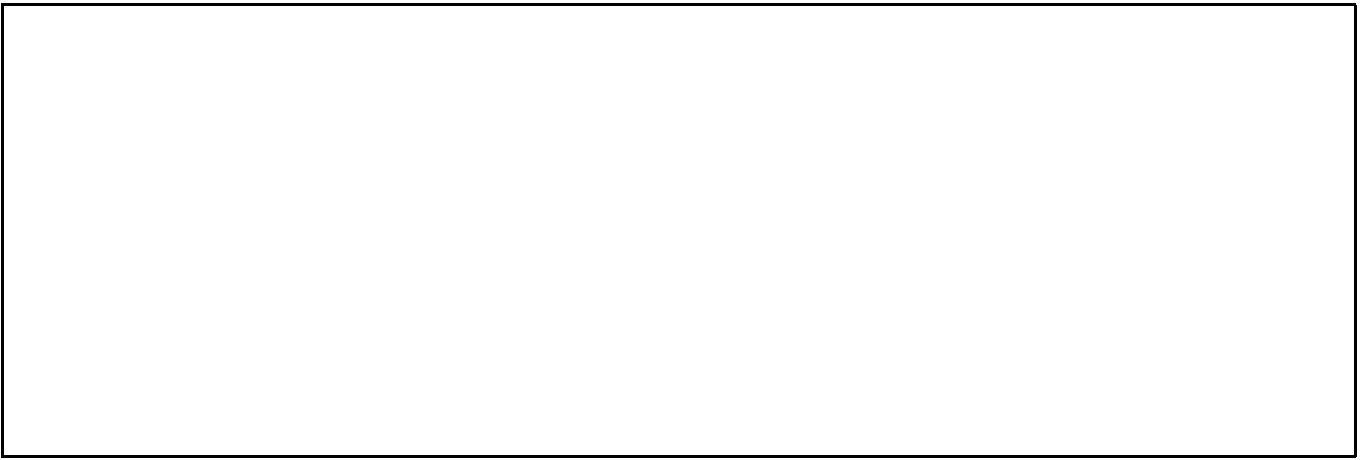

\picplace{6cm}
\caption{Time evolution of the gas clouds, same parameters as those in Fig. 3 were 
used except that the variation of $\Sigma$ was tracked at each time step. The time 
epochs for a, b and c are 50 Myr, 400 Myr and 1.0 Gyr, respectively.}
\end{figure*}

\section{Discussion}

\subsection{Validity}
The results obtained in Fig 3. are
conditional on the absence of other complicated processes during
the cloud-cloud collision, e.g., the formation of giant molecular clouds and 
subsequent instabilities, star formation
or massive star bursts, and such follow-up processes as
heating by stars, stellar wind, protostellar jets and poststellar 
explosions.
Indeed, such processes could be triggered when 
the surface density of gas clouds becomes high enough (Elmegreen 1994), 
and the motions and thermodynamic state of the gas clouds could be affected
 (Krugel \& Tutukov 1993).
Obviously, our present simulation code will not be valid in such case.

We give a very simplified example. Beckman et al ( 1991 ) suggest that 
starburst activities around the
nucleus of NGC 4736 can produce an outflow motion of $\sim 30 \kms$, as
observed by van der Kruit ( 1976 ), 
 Thus we add radial motions ($\sim 30 \kms$) to some of the gas clouds 
 inside the ILR when the surface density $\Sigma$ is higher than
a critical value 
$\Sigma_{\rm c} = \alpha \frac{\kappa c}{3.36 G} $ 
(Toomre 1964, Cowie 1981, Kennicutt 1989).
This might be a way to understand the possible influence of star formation
on the distribution 
of gas clouds in the ILR region.  The simulation obtained in this case is 
shown in Fig 4. It is interesting to find that the inner ring remains 
stable at the ILR location of about 1.1 $-$ 1.2 kpc even after 
3.6 Gyr. Although this approach to simulating the effect of
star formation is very crude, it does give a clue that the debate on
the origin of the ring structures in NGC 4736 mentioned in Sect. 1
could be reconciled in that both nuclear explosion, an event probably
similar to massive starburst, and orbital resonance are needed for keeping
the inner ring stable for a long period of time. More work on this matter
is under consideration.

\subsection{Modelling of the cloud-cloud interactions}
In comparison with others, and beyond the potential adopted for NGC 4736,
we have modified both the orbit integrator and the modelling of the 
cloud-cloud interactions. To see
whether the latter is the basic factor improving our results, as suggested by
Lia Athanassoula ( 1995, private communication ), we have repeated one of
Schwarz's simulations ( 1984 ), using exactly the same potential
as he, but our model for cloud-cloud interactions. The results are shown in
Fig 5 with the viscous coefficient $\alpha=0.0001$.
It is interesting to note that the depopulation of the 
$L_{\rm 4},L_{\rm 5}$ Lagrangian points is much more rapid than Schwarz's 
results, shown in his Fig 6 ( 1984 ). The outer ring forms after four bar
rotations, and becomes almost circular and stable after six bar rotations. 
The same structures formed in Schwarz's simulation after twenty bar rotations.
It appears that the basic difference between Schwarz's simulation and ours
is the time evolution and the time scale for stabilization.
 Thus, we believe that the 
different modelling of the cloud-cloud interactions might be the dominant
 factor responsible for the improvement in our results.

\subsection{The approximation}
An important approximation in our numerical experiments was to keep
the surface density, $\Sigma$, in eqn.(8) constant during the 
time evolution of gas clouds. Apparently, it is far from being the case, as 
the simulation indicated.

We have examined the effect of this approximation. Instead of keeping constant
surface density, $\Sigma_{0}$, a more time-consuming simulation has been
performed,
in which the variation of surface density was tracked for each step.
One such simulation  is shown in Fig 6 . As compared with those in Fig. 3,
one can 
find major differences in the inner ring, which moves closer to the nucleus and
gathers many more gas clouds.
On the other hand, the approximation has little effect on the outer ring 
structure.

Though tracking the variation of $\Sigma$ is more realistic,
that more gas clouds assemble near the galactic nucleus may not be true.
As pointed out in Sect.5.1, when the gas density exceeds the critical
value $\Sigma_{\rm c}$,
the process of star formation or even starburst must be considered instead.
The resulting superwind might prohibit the gas clouds from moving inward, 
as our simple simulation in Fig 4 indicated. So, it might not be reasonable
to track the variation of $\Sigma$ without considering the processes of 
intense star formation and the subsequent superwind. In other words, the 
approximation made in the present simulation might not be too bad.

\section{Conclusions}

In this paper, we present a new "sticky particle" model of NGC 4736, which has 
three improvements over previous models for this galaxy:
(1) The orbit integrator; (2) the potential adopted for the galaxy; and (3)
 the different modelling of cloud-cloud interactions.  
The main advantages are as follows: (1) The numerical 
experiments show that both inner and outer rings in NGC 4736 
are stable structures located at the ILR and OLR, respectively, 
more in agreement with the observations. The 
simulations overcome the drawbacks described in Sect. 1;  
(2) The symplectic algorithm has an advantage over the usual leapfrog 
method in the time step taken, which could save a lot of computational time during 
the simulations; (3) The major improvement is due to the different
modelling of the cloud-cloud interactions with the viscous force as suggested
in Sect. 5.2.

Our very simplified simulation gives us a clue that both nuclear
explosions caused by starburst activities and orbitl resonance might
be needed to keep the inner ring stable. A more realistic way to introduce 
the process of star formation as well as the cloud-cloud collision into our 
simulation is under consideration.

\begin{acknowledgements}
We would like to thank the referee, Dr. E. Athanassoula,
for her careful reading our manuscript and her instructive suggestions, 
that significantly improve our analyses in this paper.
We would also like to thank the anonymous language corrector for improving
and correcting our English presentation.
This research was supported by grants from National Science and 
Technology Commission and National Natural Science Foundation of China.
\end{acknowledgements}

\end{document}